\documentclass[aps,pre,twocolumn,showpacs,showkeys,superscriptaddress,longbibliography]{revtex4-1}

\usepackage{amsmath}
\usepackage{graphicx}
\include{MyMc}

\newif\ifPRL
\PRLfalse
\ifPRL
\newcommand{\mysec}[1]{} 
\else
\newcommand{\mysec}[1]{\section{#1}} 
\fi

\begin{document}
\title{Tunable $\varphi$ Josephson Junction ratchet}

\author{R. Menditto}
\affiliation{%
  Physikalisches Institut and Center for Quantum Science in LISA$^+$,
  Universit\"at T\"ubingen, Auf der Morgenstelle 14, D-72076 T\"ubingen, Germany
}

\author{H. Sickinger}
\affiliation{%
  Physikalisches Institut and Center for Quantum Science in LISA$^+$,
  Universit\"at T\"ubingen, Auf der Morgenstelle 14, D-72076 T\"ubingen, Germany
}

\author{M. Weides}
\affiliation{%
 Physikalisches Institut, Karlsruher Institut f\"ur Technologie,
 D-76128 Karlsruhe, Germany
}

\author{H. Kohlstedt}
\affiliation{%
  Nanoelektronik, Technische Fakult\"at,
  Christian-Albrechts-Universit\"at zu Kiel,
  D-24143 Kiel, Germany
}

\author{D. Koelle}
\author{R. Kleiner}
\author{E. Goldobin}
\email{gold@uni-tuebingen.de}
\affiliation{%
  Physikalisches Institut and Center for Quantum Science in LISA$^+$,
  Universit\"at T\"ubingen, Auf der Morgenstelle 14, D-72076 T\"ubingen, Germany
}

\date{%
  \today
}

\begin{abstract}
  We demonstrate experimentally the operation of a deterministic Josephson ratchet with tunable asymmetry. The ratchet is based on a $\varphi$ Josephson junction with a ferromagnetic barrier operating in the underdamped regime. The system is probed also under the action of an additional dc current, which acts as a counter force trying to stop the ratchet. Under these conditions the ratchet works against the counter force, thus producing a non-zero output power. Finally, we estimate the efficiency of the $\varphi$ Josephson junction ratchet.

\end{abstract}



\maketitle

\mysec{Introduction}
\nocite{Reimann:2002:BrownianMotors,Linke02:APA:SpecRatchets,Haenggi:2005:BrownMotors,Haenggi:2009:ArtBrownMotors}
\nocite{Juelicher:1997:MMM}
\nocite{Falo:1999:JJA-Ratchet,Trias:2000:JJA-Ratchet,Goldobin:2001:RatchetT,Carapella:RatchetT:2001,Carapella:RatchetE:2001,Carapella:2002:JVR-HighFreq,Lee:2003:JJA-Ratchet,Ustinov:2004:BiHarmDriverRatchet,Beck:2005:RatchetE,Wang:2009:IJJ-Ratchet,Knufinke:2012:JVR-loaded}
\nocite{Weiss:Exp-SQUID-Rats,Sterck:2002:SQUID:Ratchet,Sterck:2005:3JJ-SQUID:RockRatchet,Sterck:2009:3JJ-SQUID:StochasticRatchet,Spiechowicz:2014:JJ-SQUID:finite_capacitance,Spiechowicz:2015:Jphase:Diffusion,Spiechowicz:2015:Efficiency:SQUID}
\nocite{Villegas:2003:AVR,Savelev:2002:AVR}

Ratchets or Brownian motors attracted a lot of interest in the last decades \cite{Juelicher:1997:MMM,Falo:1999:JJA-Ratchet,Trias:2000:JJA-Ratchet,Weiss:Exp-SQUID-Rats,Carapella:RatchetE:2001,Carapella:RatchetT:2001,Goldobin:2001:RatchetT,Carapella:2002:JVR-HighFreq,Linke02:APA:SpecRatchets,Reimann:2002:BrownianMotors,Savelev:2002:AVR,Sterck:2002:SQUID:Ratchet,Lee:2003:JJA-Ratchet,Villegas:2003:AVR,Ustinov:2004:BiHarmDriverRatchet,Beck:2005:RatchetE,Haenggi:2005:BrownMotors,Sterck:2005:3JJ-SQUID:RockRatchet,Haenggi:2009:ArtBrownMotors,Sterck:2009:3JJ-SQUID:StochasticRatchet,Wang:2009:IJJ-Ratchet,Knufinke:2012:JVR-loaded,Spiechowicz:2014:JJ-SQUID:finite_capacitance,Spiechowicz:2015:Enhancement:Efficiency,Spiechowicz:2015:Diff:Anomalies,Spiechowicz:2015:Jphase:Diffusion,Spiechowicz:2015:Efficiency:SQUID,Magnasco:1993:Forced-Thermal-Ratchets,Bartussek:1994:RockedThermRatchet,Doering:1994:NoneqFluctIndTransp,Jung:1996:IntertiaRatchets,Kula:1998:Ratchet:NoiseCtrlTransp,Sarmiento:1999:Ratchets:DetTransp,Barbi:2000:UdampDetRat:PL&CR,Mateos:2000:UdampDetRat:CR,Kostur2001:BrownRat:MultiCR,Borromeo:2002:DetRat}. Apart from answering some fundamental questions, they can be immediately employed for the extraction of work out of nonequilibrium thermal fluctuations, for rectification of deterministic signals, or for particle separation \cite{Reimann:2002:BrownianMotors,Linke02:APA:SpecRatchets,Haenggi:2005:BrownMotors,Haenggi:2009:ArtBrownMotors}. Apart from the ratchets existing in nature \cite{Juelicher:1997:MMM}, there are plenty of artificial ratchet implementations, in particular, based on nanostructured superconductors: Josephson vortex ratchets \cite{Falo:1999:JJA-Ratchet,Trias:2000:JJA-Ratchet,Goldobin:2001:RatchetT,Carapella:RatchetT:2001,Carapella:RatchetE:2001,Carapella:2002:JVR-HighFreq,Lee:2003:JJA-Ratchet,Ustinov:2004:BiHarmDriverRatchet,Beck:2005:RatchetE,Wang:2009:IJJ-Ratchet,Knufinke:2012:JVR-loaded}, SQUID ratchets \cite{Weiss:Exp-SQUID-Rats,Sterck:2002:SQUID:Ratchet,Sterck:2005:3JJ-SQUID:RockRatchet,Sterck:2009:3JJ-SQUID:StochasticRatchet,Spiechowicz:2014:JJ-SQUID:finite_capacitance,Spiechowicz:2015:Jphase:Diffusion,Spiechowicz:2015:Efficiency:SQUID} and Abrikosov vortex ratchets \cite{Villegas:2003:AVR,Savelev:2002:AVR}.

A huge number of theoretical works \cite{Magnasco:1993:Forced-Thermal-Ratchets,Bartussek:1994:RockedThermRatchet,Doering:1994:NoneqFluctIndTransp,Jung:1996:IntertiaRatchets,Kula:1998:Ratchet:NoiseCtrlTransp,Sarmiento:1999:Ratchets:DetTransp,Barbi:2000:UdampDetRat:PL&CR,Mateos:2000:UdampDetRat:CR,Kostur2001:BrownRat:MultiCR,Borromeo:2002:DetRat} published more than a decade ago, were devoted to a paradigmatic system --- a point-like particle moving in a 1D periodic potential without reflection symmetry under the action of a deterministic or random force with zero time average. To create such a system using a Josephson junction (JJ), one recalls that the Josephson phase $\phi$ can be considered as the coordinate of a fictitious particle moving in a $2\pi$-periodic Josephson potential energy profile $U(\phi)$. The ratchet's driving force is the bias current. However, the Josephson potential $U(\phi)$ in most types of known JJs is reflection symmetric and its shape is hardly controllable. Thus, during many years there was no possibility to create a Josephson junction ratchet, which would be as simple as the paradigmatic examples discussed in the literature and check experimentally all the predictions. Researchers, however, were able to demonstrate more complex Josephson ratchets (with more than one JJ or with extended JJ), such as asymmetric SQUID ratchets \cite{Sterck:2002:SQUID:Ratchet, Sterck:2005:3JJ-SQUID:RockRatchet,Sterck:2009:3JJ-SQUID:StochasticRatchet,Spiechowicz:2014:JJ-SQUID:finite_capacitance} or Josephson vortex ratchets \cite{Trias:2000:JJA-Ratchet,Carapella:RatchetE:2001,Beck:2005:RatchetE,Knufinke:2012:JVR-loaded}. The physics of such devices is more complicated and they are not as reliable as the generic ratchet.

\begin{figure}[tb]
\centering
  \includegraphics[scale=0.35]{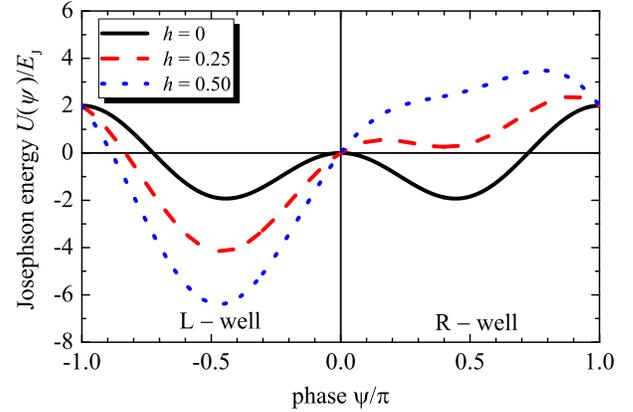}
  \caption{(color online)
    Josephson energy $U(\psi)$ tuned by an applied magnetic field $h$. Note that at any $h$ the $U(\psi)$ profile remains $2\pi$ periodic.
  }
  \label{Fig:Pot}
\end{figure}

Current progress in JJs allows us to solve this long standing problem. Recently, our group suggested \cite{Goldobin:2011:0-pi:H-tunable-CPR} and demonstrated \cite{Sickinger:2012:varphiExp} a $\varphi$ JJ with a magnetic-field-tunable Josephson energy profile. In particular, in the case of short junctions, the Josephson energy can be written in a simple analytical form as
\begin{equation}
  U(\psi) = 1-\cos(\psi) + \frac{\Gamma_0}{4}[1-\cos(2\psi)] + \Gamma_h h \sin(\psi)
  , \label{Eq:U(psi)}
\end{equation}
where $\psi=\av{\phi(x)}$ is the average Josephson phase across the JJ. The constants $\Gamma_0<0$ and $\Gamma_h$ are related to the geometrical and electrical parameters of the JJ, and $h$ is the normalized magnetic field \cite{Goldobin:2011:0-pi:H-tunable-CPR,Lipman:2014:tunable-CPR}. For longer JJs, the $U(\psi)$ profile deviates from the analytical form given by Eq.~\eqref{Eq:U(psi)}, but can be calculated numerically. In any case, the following common behavior of the $\varphi$ JJ is observed: at zero magnetic field $h=0$ the Josephson energy $U(\psi)$ is reflection symmetric (see Fig.~\ref{Fig:Pot}); at $h\neq0$ it becomes asymmetric due to the presence of both $\cos(2\psi)$ and $\sin(\psi)$ terms in Eq.~\eqref{Eq:U(psi)}. Thus, one is able not only to construct a ratchet closely mimicking the paradigmatic example, but also tune its asymmetry during experiments by changing $h$, \eg, switch it on, off, reverse its sign, etc. This is an extremely useful feature from a practical point of view as it allows to compare the transport/rectification with and without asymmetry and explore and optimize the ratchet performance by tuning the asymmetry of $U(\psi)$.

\ifPRL\else
The paper is organized as follows. In Sec. \ref{sec:Experimental results} we describe the sample design and present the experimental results of the ratchet operation in the underdamped regime. Sec. \ref{sec:Conclusions} concludes the paper.
\fi


\mysec{Experimental results}
\label{sec:Experimental results}

We used superconductor-insulator-ferromagnet-superconductor (SIFS) Josephson junctions that are fabricated as Nb$|$Al-Al$_2$O$_3$$|$Ni$_{0.6}$Cu$_{0.4}$$|$Nb multilayers \cite{Weides:2007:JJ:TaylorBarrier, Weides:2010:SIFS-jc1jc2:Ic(H)}. They consist of two segments: the first is a 0 segment of length $L_0$ with the thickness of the ferromagnetic layer $d_{F,0}$ and the critical current density $j_{c,0}>0$. The second is a $\pi$ segment of the length $L_\pi$ with the thickness of the ferromagnetic layer $d_{F,\pi}$ and $j_{c,\pi}<0$. Such a JJ as a whole behaves \cite{Goldobin:2011:0-pi:H-tunable-CPR,Lipman:2014:tunable-CPR} as a $\varphi$ JJ with the average phase $\psi=\av{\phi(x)}$ and the Josephson energy $U(\psi)$ qualitatively similar to the one given by Eq.~\eqref{Eq:U(psi)}. The exact $U(\psi)$ profile can be calculated only numerically \cite{Sickinger:2012:varphiExp}. In any case it is important that at bias current $I=0$ and magnetic field $h=0$,  $U(\psi)$ is a reflection symmetric $2\pi$ periodic double well potential with the minima of the wells at $\psi=\pm\varphi+2\pi n$, see Fig.~\ref{Fig:Pot}. At $h=0$ the wells are degenerate, while for $h \neq 0$ the degeneracy is removed \cite{Goldobin:2011:0-pi:H-tunable-CPR,Lipman:2014:tunable-CPR,Sickinger:2012:varphiExp}.

\begin{figure}[!tb]
  \includegraphics[scale=0.35]{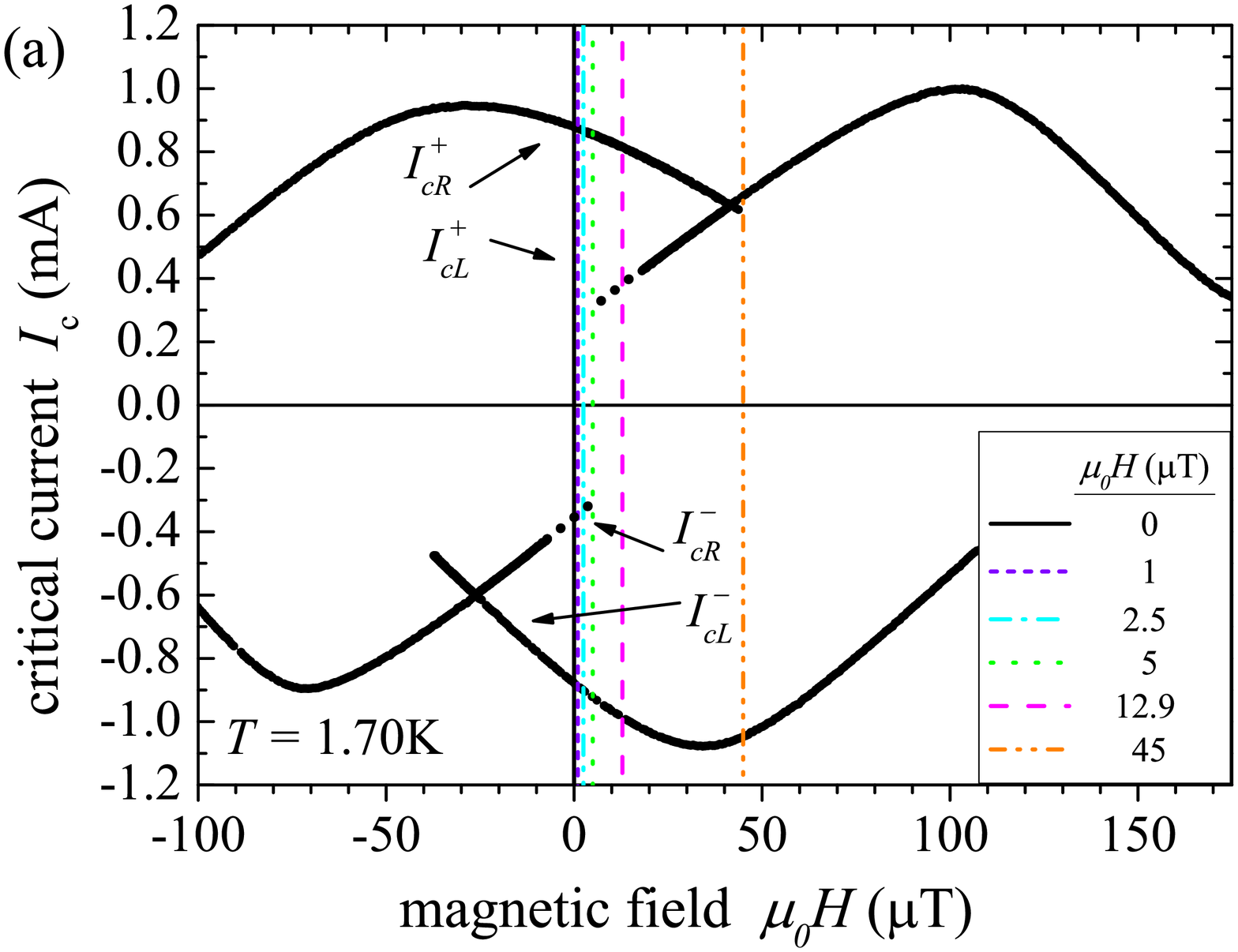}
  \vspace{5mm}
  \includegraphics[scale=0.35]{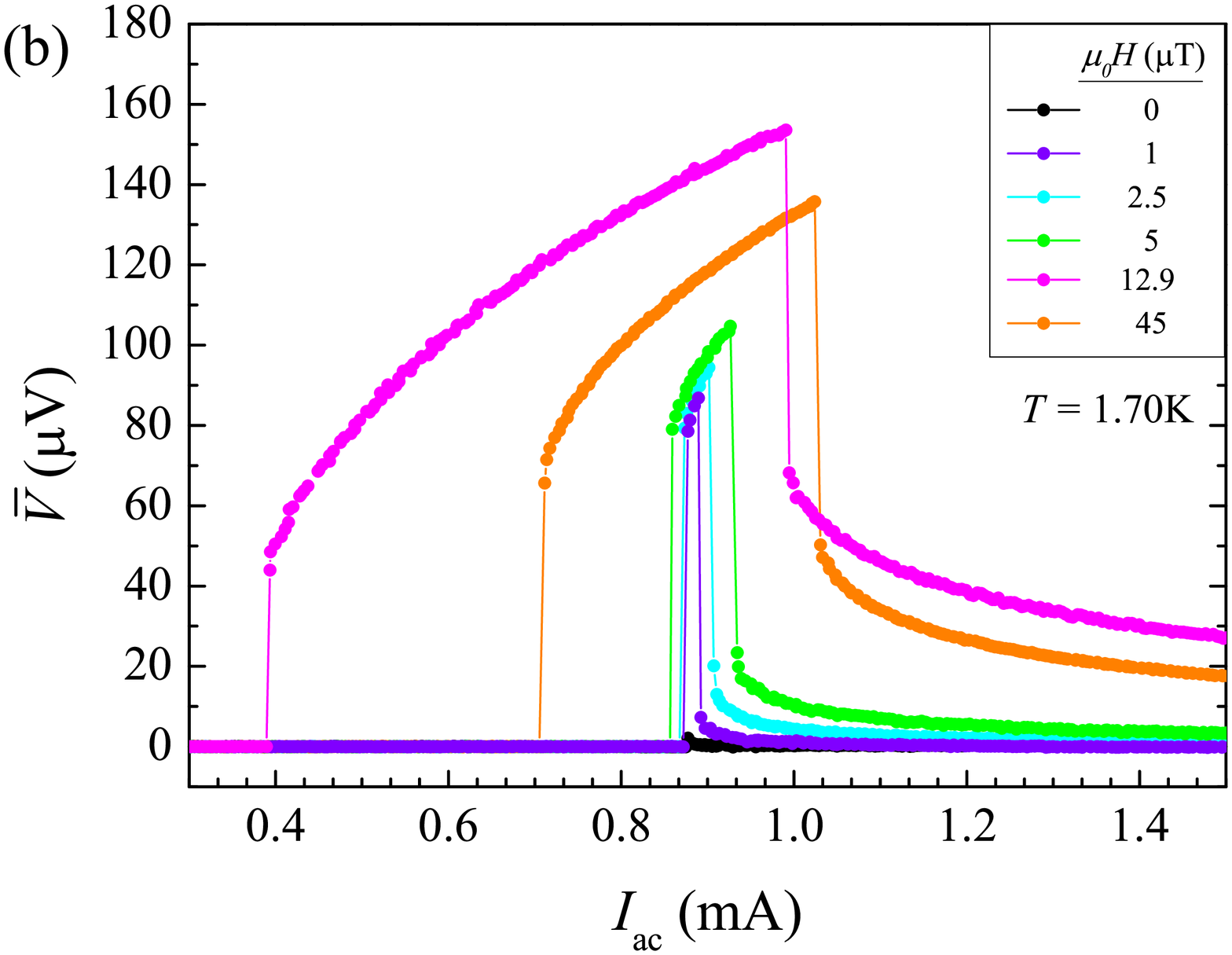}
  \caption{(Color online)
    (a) $I_c(H)$ curve at $T=1.70\units{K}$. Vertical lines in (a) correspond to the values of $\mu _{0}H$, for which different rectification curves $\mean{V}(I_{ac})$ in (b) are measured.
  }
  \label{Fig:IcH}
\end{figure}

As demonstrated in our previous works \cite{Goldobin:CPR:2ndHarm,Sickinger:2012:varphiExp} a typical property of a $\varphi$ JJ is to have two critical current branches, denoted here as $I_{c,L}^+(H)$ and $I_{c,R}^+(H)$, measured for increasing\cite{NoteIc} (superscript ``$+$'') bias current and two branches denoted as $I_{c,L}^-(H)$ and $I_{c,R}^-(H)$ for decreasing\cite{NoteIc} (superscript ``$-$'') bias current, see Fig.~\ref{Fig:IcH}(a). These two currents correspond to the escape of the phase out of the left ``L'' and the right ``R'' wells of $U(\psi)$, see Fig.~\ref{Fig:Pot}. The smaller (by amplitude) of the two critical currents (at a given $H$) can be observed only for low enough damping. For higher damping, upon the escape from, \eg, the L-well, the phase can be retrapped in the R-well. Consequently one will observe $I_{c,R}^+$ when the phase will later on escape from the R-well instead of $I_{c,L}^+$. In general, the damping in SIFS JJs is strongly temperature dependent and reduces for lower temperatures \cite{Pfeiffer:2008:SIFS-0-pi:HIZFS}. For our samples we estimated $T=3.60\units{K}$ as the crossover temperature between the high and low damping regime (in a sense of observing both $I_{c,LR}$'s).

Our measurements were performed in a $^3$He cryostat, equipped with a multi-layer magnetic shielding. All electrical connections (wires) going to/from the sample have been filtered both at room-temperature and at cryogenic temperatures. The magnetic field was applied by a coil with $\mu _{0}H=\eta \cdot I_{\textrm{coil}}$ with coil factor $\eta \sim 5 \units{\mu T/ mA}$.

\begin{figure*}[]
   \includegraphics[scale=0.35]{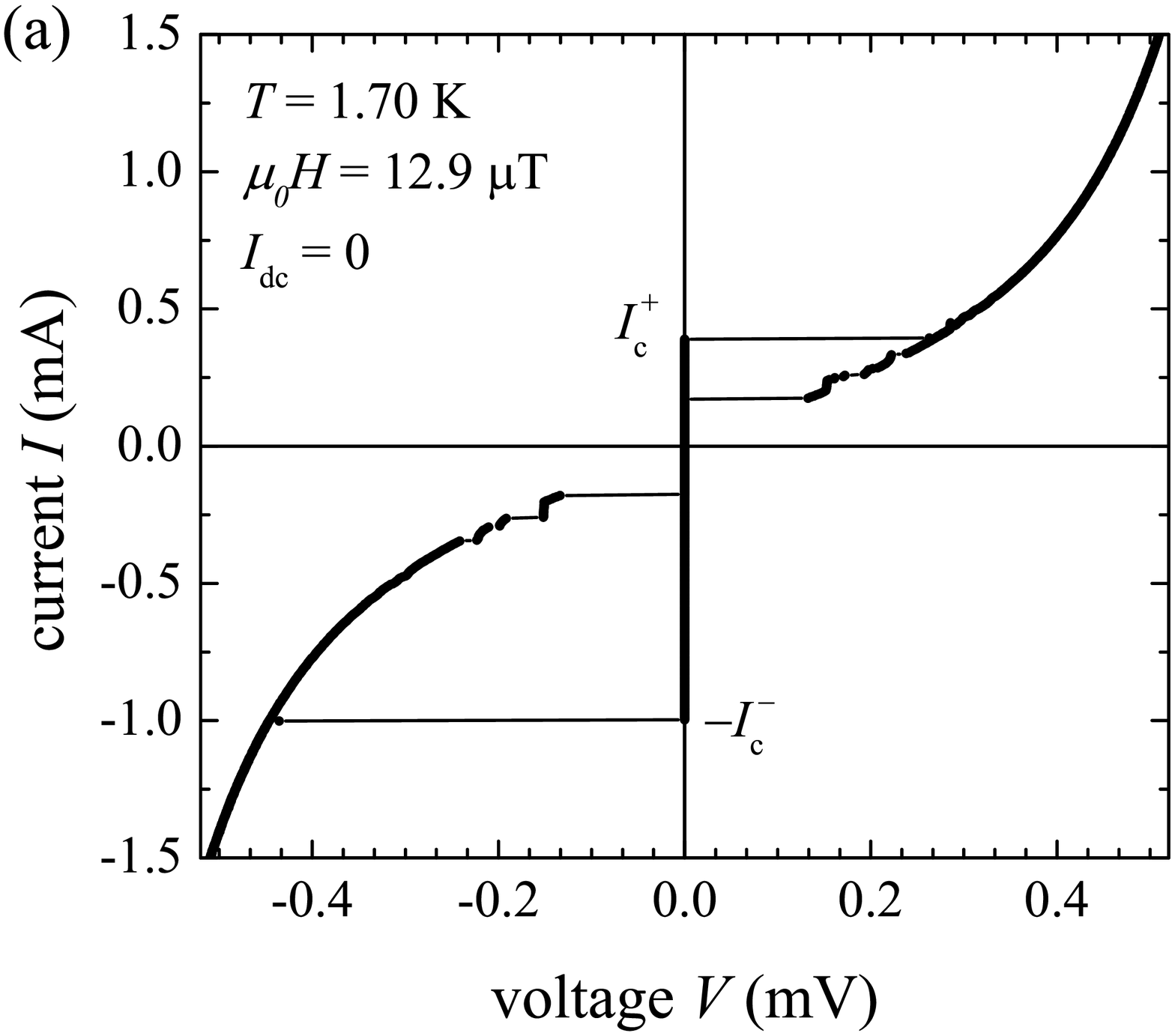}
   \includegraphics[scale=0.35]{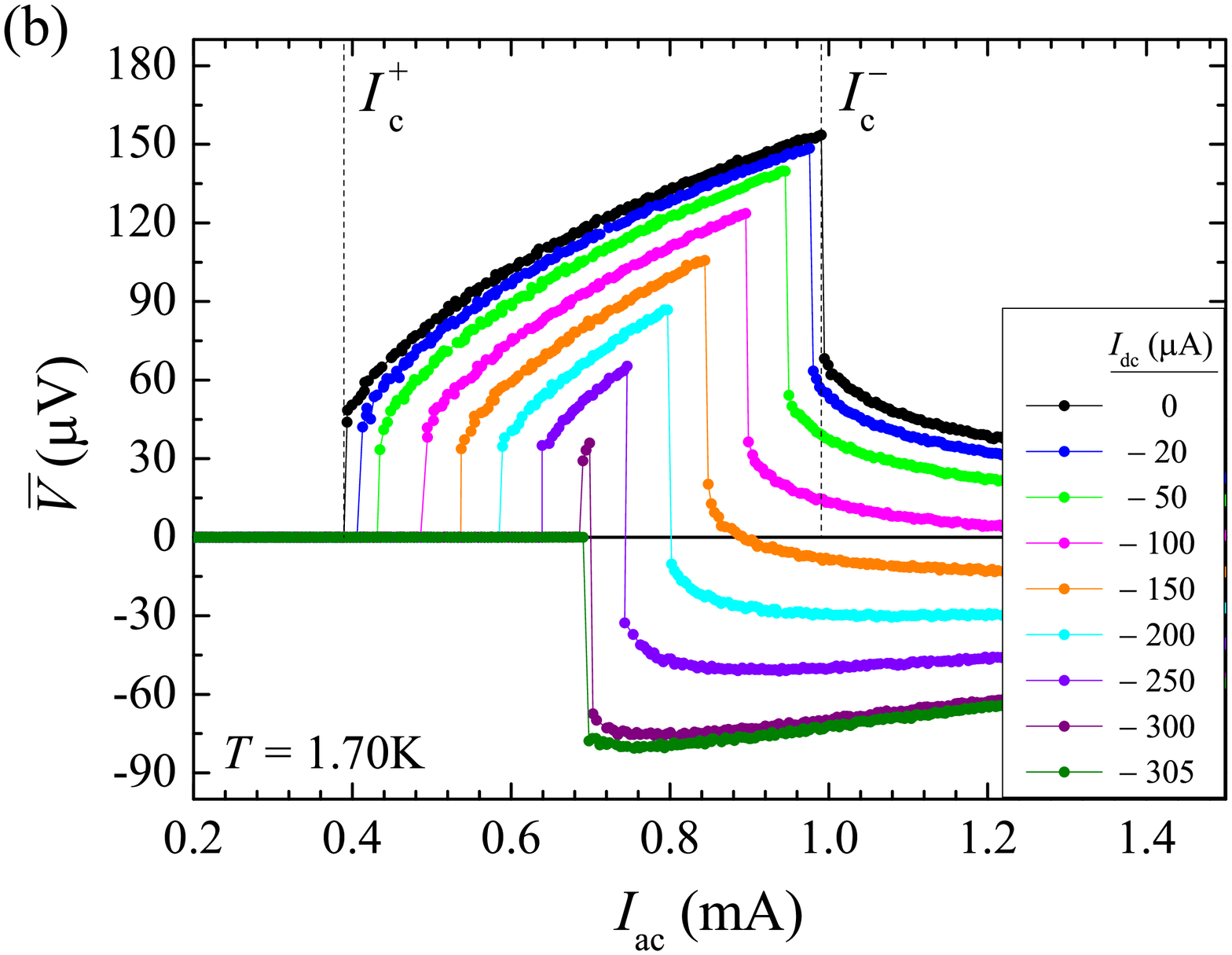}
   \caption{(Color online)
     $\varphi$ JJ at $T=1.70 \units{K}$ and $\mu _{0}H= 12.9 \units{\mu T}$. (a) Current-voltage characteristics and (b) rectification curves for different amplitudes of the counter force $I_{dc}$. In (a) $I_c^+\approx389\units{\mu A}$, $I_r^+\approx171\units{\mu A}$ and $I_c^-\approx997\units{\mu A}$.}
  \label{Fig:stop}
\end{figure*}

The dependence of the critical current $I_c$ on the externally applied magnetic field $H$ at $T=1.70\units{K}$ is shown in Fig.~\ref{Fig:IcH}(a). The existence of two critical current branches $I_{c,L}^\pm$ and $I_{c,R}^\pm$ as well as the crossing of the branches, typical of a $\varphi$ JJ, is observed. Two $I_c$'s are well visible for $-37\units{\mu T} \lesssim \mu _{0}H \lesssim-7\units{\mu T}$ and $18\units{\mu T} \lesssim \mu _{0}H \lesssim44\units{\mu T}$. However, for $-7\units{\mu T} \lesssim \mu _{0}H \lesssim18\units{\mu T}$ for this particular JJ and $T$, the $I_{c,L}^+$ and $I_{c,R}^-$ branches are semi-stable (do not always appear), which is indicated by the dots continuing these branches, see Fig.~\ref{Fig:IcH}(a). The traceability of the lower (by absolute value) $I_c(H)$ branches in experiment also depends on the bias current sweep sequence, \ie, depends on the well, L or R, in which the phase is trapped initially. The sweep sequences are rather different for measurements of $I_c(H)$ and rectification curves, see Figs.~\ref{Fig:IcH}(b) and \ref{Fig:stop}(b).
By applying a magnetic field one can change the asymmetry between the wells of the Josephson potential energy $U(\psi)$ and create an asymmetric periodic potential required for a ratchet operation, \cf Fig.~\ref{Fig:Pot}.


Here we present the results obtained in the underdamped regime at $T=1.70\units{K}$, where the rectification operation is strong and rectification curves $\mean{V}(I_{ac})$ appear free from extra structures due to the presence of (half-integer zero field) steps on the current-voltage characteristics (IVCs) (see the steps, \eg, in Fig.4 of Ref.~\onlinecite{Sickinger:2012:varphiExp}).


In our experiment we measure the rectification curves $\mean{V}(I_{ac})$, \ie, the average voltage \vs the amplitude of applied ac current. For this we apply a periodic bias current $I(t)=I_{ac}\sin(2\pi f t)$ with the frequency $f=10\units{Hz}$ and the update rate of $10000\units{pts/s}$ (period $T=100\units{ms}$, $1000\units{pts/period}$) and we measure the voltage $1000$ times with the sampling rate $10000\units{samples/second}$, \ie, exactly during one period, starting at an arbitrary moment of time defined by delays in hardware and software. Then the collected samples are averaged to obtain $\mean{V}$ at given $I_{ac}$. The sign of $\mean{V}$ indicates the direction of motion of the phase in the Josephson potential. In the following we discuss the case of $\mean{V}>0$, \ie, $I_c^+<|I_c^-|$, where $I_c^+$ or $I_c^-$ mean the \emph{relevant}, L or R, $I_c^\pm(H)$ branch, see below for details. The opposite situation ($\mean{V}<0$) is similar.

For small $I_{ac}$ the current is so small, that it does not exceed $I_c^+$, so that the phase remains pinned in the well and $\mean{V}=0$. If $I_{ac}$ becomes larger, \ie  $I_{c}^+<I_{ac}<|I_{c}^-|$, the voltage becomes $\mean{V}\neq 0$, because for $I_{ac}>I_c^+$ the JJ jumps to the resistive branch. In the underdamped regime, due to the hysteresis on the IVC, the voltage $\mean{V}$ jumps to a finite value at the beginning of the rectification region. Then for $I_{ac}>|I_c^-|$ the voltage $\mean{V}$ decreases because the junction also picks up some negative voltage during the negative semiperiod.
In Fig.~\ref{Fig:IcH}(b), the $\mean{V}(I_{ac})$ curves are shown for different values of the magnetic field $H$, \ie for different asymmetries of the energy potential $U(\psi)$. First, at $\mu_0 H=0$, the rectification is absent ($\mean{V}$=0), for any amplitude $I_{ac}$ of the driving current $I(t)$. In the absence of external field the energy potential is reflection symmetric, therefore no rectification is expected. As soon as the field is applied to the JJ, the reflection symmetry of the potential is broken and unidirectional motion of the phase occurs, see Fig.~\ref{Fig:IcH}(b). The width of the rectification window changes with the applied magnetic field, reflecting the change in the asymmetry of the energy potential and, therefore, $I_c^\pm$.
For $|\mu _{0}H|<10\units{\mu T}$ we see rather narrow rectification windows due to the small difference in $I_{c,R}^+(H)$ and $|I_{c,L}^-(H)|$. For $10\units{\mu T}<|\mu_0 H|<40\units{\mu T}$ the $I_{c,L}^+$ instead of $I_{c,R}^+$ comes into play. As a result the rectification window increases substantially, see Fig.~\ref{Fig:IcH}(b). For even larger $|\mu_0 H|\lesssim40\units{\mu T}$ the rectification window narrows somewhat because the difference between $|I_{c,L}^+$ and $|I_{c,L}^-|$ decreases, see Fig.~\ref{Fig:IcH}(a).



Up to now the ratchet shows operation in the idle regime ($I_{dc}=0$). We now apply an additional dc bias current $I_{dc}$ (counter force) to the ratchet, which tries to stop the ratchet or even move the phase in the direction opposite to the rectification direction. If the ratchet is able to overcome the counter force $I_{dc}$, it produces a mean output power $\mean{P}_\mathrm{out}=I_{dc}\mean{V}<0$ (\ie  the work is done by the ratchet on the current source). Furthermore one can calculate the efficiency, given by $\eta = - \mean{P}_\mathrm{out}/\mean{P}_\mathrm{in}$, where $\mean{P}_\mathrm{in}$ is the mean input power.

To demonstrate the operation of the ratchet against the counter force, we have chosen the value of $\mu _{0}H=12.9 \units{\mu T}$, where the rectification window is largest. Fig.~\ref{Fig:stop}(a) shows the IVC of the device for this value of $H$. Here the relevant  $I_c^+\equiv I_{c,L}^+(H)$ and $I_c^-\equiv I_{c,L}^-(H)<0$, \ie, the pinning/depinning game takes place in the L-well, which becomes deeper at $H>0$, while the R-well becomes more shallow and may even disappaer.

The stopping force $I_\mathrm{stop}(I_{ac})$ is defined as the current $I_{dc}$ at which $\mean{V}$ (within the rectification window in the idle regime) vanishes or changes sign at a given $I_{ac}$. We measured many rectification curves $\mean{V}(I_{ac})$, each time increasing the amplitude of the dc current. Since we have a positive rectification, $\mean{V}>0$, the counter force $I_{dc}<0$ should be negative.
The results are shown in Fig.~\ref{Fig:stop}(b). Starting from the curve with $I_{dc}=0$, one can see that by increasing the absolute value of $I_{dc}$ the rectification window narrows, indicating that the additional bias actually stops the ratchet at the regions where the ratchet was not strong enough (edges of the idle rectification window).
Note that the shrinkage is symmetric relative to the center of the rectification window, and this is due to the fact that the constant bias shifts up all the currents of the IVC.
From these measurements we see that the at $I_{dc}^\mathrm{off}\approx-307\units{\mu A}$ the rectification window closes completely and the ratchet operation stops fully. The theoretical value can be calculated using our parameters (see the caption of Fig.~\ref{Fig:stop}) as \cite{Goldobin:model-ratchet} $I_{dc}^\mathrm{off}= (I_c^{+} - |I_c^{-}|)/2=-304\units{\mu A}$, which is a rather exact coincidence with the experimental value. According to the theory \cite{Goldobin:model-ratchet} the full-stop force $I_{dc}^\mathrm{off}$ depends \emph{only} on $I_{c}^{+}$ and $I_{c}^{-}$ but not on the shape of the IVC.

For given $I_{dc}$, the maximum efficiency is always reached at $I_{ac}=J_{c}^+ \equiv I_{c}^{+} - I_{dc}$, \ie, in the beginning of the rectification window and is given by \cite{Goldobin:model-ratchet}
\begin{equation}
  \eta_\mathrm{max} = \frac{-2I_{dc}\left[ I_{dc}\arccos\left(\frac{J_r^+}{J_c^+}\right) + \sqrt{{J_c^+}^2-{J_r^+}^2} \right]}
    {{J_c^+}^2\arccos\left(\frac{J_r^+}{J_c^+}\right) + \sqrt{{J_c^+}^2 - {J_r^+}^2}(J_r^+ + 2I_{dc})}
  ,\label{Eq:eta_step_max}
\end{equation}
where $J_{r}^+=I_{r}^{+} - I_{dc}$, with $I_{r}^{+}$ the return current from the resistive branch.

Using Eq.~\eqref{Eq:eta_step_max} and our parameters (see the caption of Fig.~\ref{Fig:stop}) we may plot the dependence $\eta_\mathrm{max}(I_{dc})$ given by Eq.~\eqref{Eq:eta_step_max}. This dependence (not shown) smoothly grows with $|I_{dc}|$. The maximum value of $|I_{dc}|$ that makes sense is $I_{dc}^\mathrm{off}$ measured and calculated above. At this $I_{dc}$ the rectification window is about to close completely, but the ratchet is the most efficient with $\eta_\mathrm{max}=48\units{\%}$. This is a fairly good value, which is not much lower than the maximum efficiency of $\eta_\mathrm{max}=60\units{\%}$ observed in a specially designed vortex ratchet \cite{Knufinke:2012:JVR-loaded}. We stress here that our $\varphi$ JJ was not optimized or designed for operation as a ratchet. It is one of two samples used in the original experimental work on $\varphi$ JJs \cite{Sickinger:2012:varphiExp}.

\mysec{Conclusions}
\label{sec:Conclusions}

Although there were many theoretical studies on ratchets where the particle moves in an asymmetric periodic potential, the practical implementation of a simple paradigmatic system using a Josephson junction was missing, mainly, because the Josephson energy in conventional junctions is reflection symmetric. Here we have demonstrated that in $\varphi$ Josephson junctions this symmetry is broken and one can obtain rectification as a result of directed transport of the phase. The advantage of this system is that the asymmetry is tunable by a magnetic field $H$, so that one can clearly see the (dis)appearance of rectification as a function of $H$, as well as optimize its operation. The maximum efficiency that can be obtained with such a ratchet is rather high, considering that the parameters of the investigated junction (\eg the asymmetry of the 0 and  $\pi$ part) are not optimized for the ratchet operation.

A $\varphi$ JJ is only one example of constructing a system with desired non-trivial Josepshon energy profile $U(\psi)$. Following this general approach, one can try to design even more asymmetric ratchets that will provide a huge rectification window and, consequently, have higher full-stop current $I_{dc}^\mathrm{off}$ and higher efficiency $\eta$.

\acknowledgements
R.M. gratefully acknowledges support by the Carl Zeiss Stiftung. This work was supported by the Deutsche Forschungsgemeinschaft (DFG) via Project No. GO-1106/5, via project A5 within SFB/TRR-21, and by the EU-FP6-COST action MP1201.

\bibliography{this,SF,SFS,pi,software,JJ,LJJ,ratch}

\begin{thebibliography}{46}
\bibitem [{Not()}]{NoteIc}%
  \BibitemOpen
  \href@noop {} {}\bibinfo {note} {Here increasing and decreasing is understood
  \emph{not} by absolute value. That is $I_{c,LR}^+$ are measured when the bias
  current is swept towards positive direction (increasing), while $I_{c,LR}^-$
  are measured when the bias current is swept towards negative direction
  direction (decreasing).}\BibitemShut {Stop}%
\end{thebibliography}

\end{document}